\documentclass[aps,twocolumn,nofootinbib]{revtex4}
\usepackage[latin1]{inputenc}
\usepackage{epsfig}
\newcommand{\beq}{\begin{equation}}
\newcommand{\eeq}{\end{equation}}
\newcommand{\la}{\langle}
\newcommand{\ra}{\rangle}

\begin{document}

\title{Classical and quantum stochastic thermodynamics}

\author{Mário J. de Oliveira}
\affiliation{Universidade de São Paulo, Instituto de Física,
Rua do Matão, 1371, 05508-090 São Paulo, SP, Brazil}

\begin{abstract}

The stochastic thermodynamics provides a framework for the
description of systems that are out of thermodynamic equilibrium.
It is based on the assumption that the elementary constituents are
acted by random forces that generate a stochastic dynamics, which
is here represented by a Fokker-Planck-Kramers equation.
We emphasize the role of the irreversible probability current,
the vanishing of which characterizes the thermodynamic equilibrium
and yields a special relation between fluctuation and dissipation.
The connection to thermodynamics is obtained
by the definition of the energy function and the entropy
as well as the rate at which entropy is generated.
The extension to quantum systems is provided by a quantum
evolution equation which is a canonical quantization of 
the Fokker-Planck-Kramers equation.
An example of an irreversible systems is presented
which shows a nonequilibrium stationary state with an unceasing
production of entropy. A relationship between the fluxes and the
path integral is also presented.

\end{abstract}

\maketitle

\section{Introduction}

Thermodynamics was conceived as a discipline based on principles
and laws that refer to macroscopic quantities
such as the principles of energy conservation
and of entropy increase, which are the first and second laws
of thermodynamics. Although these two principles are valid for
systems in equilibrium as well as for systems out of equilibrium,
the initial development of the discipline lead to the establishing of
a theory of thermodynamics of system in equilibrium. This was 
possible because the energy of a system in thermodynamic equilibrium
is related functionally to the entropy, which
allows the definition of temperature. 

The derivation of the thermodynamics from the microscopic
laws of motion was the aim of the kinetic theory. 
One of its consequences was the development of
the equilibrium statistical mechanics advanced by Gibbs.
The statistical
mechanics is based on the description of system by the 
probability distribution that bears the name of Gibbs,
which for a system in contact with
a heat reservoir is proportional to $e^{-E/kT}$,
where $E$ is the energy function and $T$, the temperature.
The crucial property of the equilibrium distribution is
that the probability depends on the states of the system
only through the energy function. This property,
along with the Gibbs expression for the entropy, leads to the relation
between energy and entropy, mentioned above, which characterizes
the thermodynamic equilibrium. 

The entropy of an isolated system in equilibrium
remains invariant. But if it is not in equilibrium, its
entropy increases and the increase, in this case, is not due
to the flux of entropy because the system is isolated.
Entropy is being created spontaneously and in this sense it differs from
the energy which is a conserved quantity. 
If a system is not isolated then the variation of the entropy $S$
with time is the algebraic sum of two terms,
\beq
\frac{dS}{dt} = \Pi - \Phi,
\eeq 
where $\Pi$ is the rate in which entropy is being created,
the rate of entropy production, 
and $\Phi$ is the flux of entropy.

The production of entropy is related to irreversible processes 
occurring inside the system which are understood as processes
that are more likely to occur than their time-reverse counterparts. 
Thermodynamic equilibrium is thus characterized as the
state where a process and its time reversal are equally probable.
This characterization of equilibrium, embodied in the
stochastic thermodynamics, 
is a dynamical definition, being more comprehensible than
the static definition given above in terms of the Gibbs distribution.
The stochastic thermodynamics
\cite{schnakenberg1976,jiuli1984,mou1986,perez1994,
sekimoto1998,lebowitz1999,tome1997,jarzynski1997,mazur1999,maes2003,
crochik2005,zia2006,andrieux2006,tome2006,schmiedl2007,harris2007,
seifert2008,blythe2008,esposito2009,esposito2009a,
tome2010,broeck2010,jarzynski2011,tome2012,spinney2012,seifert2012,
santillan2013,luposchainsky2013,tome2015,oliveira2016,tome2018,
oliveira2019}
provides an approach to out of equilibrium an 
equilibrium thermodynamics that takes into account the dynamical
characterization of the irreversible processes
by assuming that a system evolves in time according
to a microscopic stochastic dynamics.

The elementary constituents
of a system are assumed to be acted by random forces
in addition to the usual deterministic forces.
As a consequence the trajectory followed by
a particle is not in general deterministic. There are many possible
trajectories that a particle may follow from a given point,
each one with a certain probability of occurrence. 
The approach we follow here uses a representation of the
dynamics in terms of the probability of the occurrence
of a state at a certain instant of time, which
is assumed to be governed by an evolution equation.

The main features of the approach that we follow here are:
(1) a stochastic dynamics which is here
represented by a Fokker-Planck-Kramers equation
\cite{kampen1981,gardiner1983,risken1984,tome2015L};
(2) the assignment of an energy function;
(3) the definition of entropy as having the same form
as the Gibbs entropy;
(4) a proper definition of the rate of entropy production.

The first part of this text is dedicated to the classical case.
In the second part we extend the results obtained in the
first part to the quantum case. In particular, we use as
the evolution equation a quantum version of the Fokker-Planck-Kramers
equation. 
In this case, the probability density is replaced
by the probability density operator, usually called density
matrix. In a third part we generalize the results for the case
of many degrees of freedom and present an example of a system
that displays a nonequilibrium stationary state with an unceasing
production of entropy.

\section{Evolution equation}

Our object of study is a system of particles that interact 
among themselves and may also be subject to external forces.
In addition each particle is acted by random forces, the origin
of which may be internal or external to system. 
The system evolves in time according to the Newton equation
of motion. Due to random forces, the trajectory is not
uniquely defined. There are many possible trajectories
starting from a given state, each one with a certain
probability.

If the system is at a given state at the initial time, one may
ask for the probability that it is at a given state at a
later time. An answer to this question is provided by
the Fokker-Planck-Kramers (FPK) equation which gives the time
evolution of the probability density $\rho(x,p,t)$.
We will focus initially on a system with just one
degree of freedom in which case $x$ is the position and $p$
is the momentum of the particle, and both quantities
constitute the state of the system. The probability that
at time $t$ the state of the system is inside $dxdp$ around
$(x,p)$ is $\rho(x,p,t)dxdp$. 

In contrast to the equilibrium statistical mechanics for which
the probability density is constant in time, here it
depends on time. If we wish that the system reaches thermal
equilibrium for long times, usually called thermalization,
then the solution of the
evolution equation for long times must be a
Gibbs equilibrium distribution, which
is characterized by depending on $(x,p)$ only through the
energy function associated to the system.

The FPK equation is given by
\beq
\frac{\partial\rho}{\partial t} =
- \frac{p}{m}\frac{\partial\rho}{\partial x}
- \frac{\partial f \rho}{\partial p}  
+ \frac{\Gamma}2 \frac{\partial^2\rho}{\partial p^2},
\label{2}
\eeq
where $m$ is the mass of the particle,
$f$ is the ordinary force acting on the particle
and $\Gamma$ is a constant that is associated to the stochastic forces.
The force $f$ is understood to be a sum of an internal force $F^c$,
considered to be a conservative force, and a dissipative force $D$,
\beq
f = F^c + D.
\eeq
The property of the dissipative force $D$ that distinguishes
it from the other forces is that it is an odd function
of the momentum.

A derivation of the FPK equation (\ref{2}) can be obtained from
a Langevin equation and can be found in reference \cite{tome2015L}. 
The Langevin equation leading to (\ref{2}) 
is the equation of motion for a particle of
mass $m$ moving along a straight line which in addition to the
ordinary force $f$ is also under the action of
a stochastic force with zero mean and variance proportional to
$\Gamma$. 

An essential property of the FPK equation, and for that matter
of any equation that governs the time evolution of
a probability distribution, is that it preserves the normalization
of $\rho$, that is,
\beq
\int \rho(x,p,t)\,dxdp = 1,
\label{3}
\eeq
for any instant of time, where the integration is performed
on the whole space of states. If $\rho$ is normalized at
the initial time, it remains normalized forever.
The demonstration of this fundamental property of the
FPK equation is given in the appendix \ref{A}.

As $F^c$ is a conservative force, we may write
$F^c = - (\partial H/\partial x)$, $p/m = (\partial H/\partial p)$,
where 
\beq
H = \frac{p^2}{2m} + V(x)
\eeq
is the energy function, which is the sum of the kinetic energy and
the potential energy $V$.
The first term and the one involving $F^c$ of the FPK equation become
\beq
- \frac{p}{m}\frac{\partial\rho}{\partial x}
- \frac{\partial F^c\rho}{\partial p}
= - \frac{\partial H}{\partial p}\frac{\partial\rho}{\partial x}
+ \frac{\partial H}{\partial x}\frac{\partial\rho}{\partial p}.
\eeq
The right-hand side of this equation is written in the abbreviated form as
\beq
\frac{\partial H}{\partial x}\frac{\partial\rho}{\partial p}
- \frac{\partial H}{\partial p}\frac{\partial\rho}{\partial x}
= \{H,\rho\},
\eeq
which is called the Poisson brackets. Replacing this result in the
equation (\ref{2}), the FPK equation acquires the form
\beq
\frac{\partial\rho}{\partial t} = \{H,\rho\}
- \frac{\partial J}{\partial p},
\label{7}
\eeq
where
\beq
J =  D\rho - \frac{\Gamma}2 \frac{\partial\rho}{\partial p}.
\label{6}
\eeq

The FPK equation (\ref{2}) can also be written in the form
\beq
\frac{\partial\rho}{\partial t} =
- \frac{\partial J_x}{\partial x} - \frac{\partial J_p}{\partial p},
\label{3aa}
\eeq
where $J_x = p/m$ and $J_p = F^c\rho + J$.
In this form, the FKP equation is understood as a 
continuity equation and $J_x$ and $J_p$ are understood
as the components of the probability current. 
This is the form that allowed us to show the property (\ref{3}),
as presented in the appendix \ref{A}.
The part $J$ of the component $J_p$ is the irreversible
probability current, which plays a fundamental role in the
present approach. We remark that without $J$
the FPK equation reduces to the Liouville equation of classical
statistical mechanics \cite{salinas2001},
\beq
\frac{\partial\rho}{\partial t} = \{H,\rho\}.
\label{12}
\eeq

\section{Thermodynamic equilibrium}

The FPK equation as it stands may or may not describe a system
that for long times will be in thermodynamic equilibrium.
For long times the solution of the FPK equation is its stationary 
solution, that is, the solution obtained by setting to zero the
right-hand side of the equation, in which case $J$ may or may not vanish.

A system in thermodynamic equilibrium may be said to be the
one described by a Gibbs distribution.
However, this is a static definition. We need here
a dynamic characterization of thermodynamic equilibrium.
This is provided by characterizing the thermodynamic equilibrium
as the stationary state such that the irreversible current vanishes.
Therefore, the equilibrium distribution $\rho_e$ obeys the
condition
\beq
D\rho_e - \frac{\Gamma}2 \frac{\partial\rho_e}{\partial p} = 0,
\label{5}
\eeq
and in addition
\beq
\{H,\rho_e\} = 0.
\eeq
The solution of this last condition leads to the result that
$\rho_e$ is a function of $H$, that is, $\rho_e$ depends on
$x$ and $p$ through the energy function $H(x,p)$. This
characterizes any Gibbs equilibrium distribution.

The condition (\ref{5}) is understood as the relation between
dissipation, described by $D$, and noise or fluctuations, described
by $\Gamma$.
That is, in equilibrium there must be a relation between
dissipation and fluctuations. We wish to describe a system
in contact with a heat reservoir at a temperature $T$, in which
case the appropriate Gibbs equilibrium distribution is
\cite{salinas2001}
\beq
\rho_e = \frac1Z e^{-\beta H},
\label{1}
\eeq
where $\beta=1/k T$ and $k$ is the Boltzmann constant.
Replacing (\ref{1}) in equation (\ref{5}), we find
\beq
D = - \gamma p,
\label{12a}
\eeq
where $\gamma$ is the constant that connects $\Gamma$ with temperature,
\beq
\Gamma = 2m\gamma k T.
\label{12b}
\eeq
Notice that $D$ is the usual type of dissipation force proportional
to the velocity.

\section{Energy, heat and entropy}

A thermodynamic system may have its energy changing in time.
The energy variation is due to the exchange of heat or work
with the environment. Here we will treat the case where
the external forces are absent so that the variation
in energy is only due to the exchange of heat. The energy $U$
of a thermodynamic system, sometimes called internal energy,
is the average of the energy function introduced above, that is,
\beq
U = \int H \rho dx dp,
\label{17}
\eeq
and may depend on time as $\rho$ depends on time.
The variation of $U$ with time is 
\beq
\frac{dU}{dt} = \int H \frac{\partial\rho}{\partial t}dxdp.
\eeq
Replacing the time derivative of $\rho$ by using the 
FPK equation in the form (\ref{7}), we find
\beq
\frac{dU}{dt} = - \int H\frac{\partial J}{\partial p} dxdp.
\label{8}
\eeq
The term involving the Poisson brackets, vanishes
after an integration by parts has been performed.
Here and in the
following, whenever we do an integration by parts, the
integrated term is assumed to vanish. As shown in the
appendix \ref{A}, this is so because
we are considering that at the limits of integration the
probability density vanishes rapidly. 
Performing the integral in (\ref{8}) by parts, we find
\beq
\frac{dU}{dt} = \int J\frac{\partial H}{\partial p} dxdp.
\eeq
The right-hand side of this equation is interpreted as the
rate at which heat is introduced into the system, or the
flux of heat $\Phi_q$,
\beq
\Phi_q = \int J\frac{\partial H}{\partial p} dxdp.
\label{8a}
\eeq
Thus we may write
\beq
\frac{dU}{dt} = \Phi_q,
\label{8b}
\eeq
an equation that may be understood as the conservation of energy.
Notice that the flux of heat $\Phi_q$ involves the irreversible
part of the probability current.

From thermodynamics, we know that heat is related to
entropy through the Clausius equation $dQ=TdS$, valid for
systems in equilibrium, where $dQ$ is the infinitesimal
heat exchanged with the system and $dS$ is the infinitesimal
increase of the entropy $S$. 
Clausius equation is equivalent to the equation
$\Phi_q=T(dS/dt)$ which could be used to define entropy.
However, this equation is of no use here because it is
valid only for system in equilibrium. The appropriate way
to define entropy of a system is to use the Gibbs form
\beq
S = - k \int\rho \ln\rho \,dxdp,
\label{9}
\eeq
which is a generalization of the Boltzmann entropy $S=k \ln W$
where $W$ is the number of accessible states. Although,
$S$ given by (\ref{9}) is usually used for system in equilibrium,
here we are assuming that this form is also appropriate for
systems out of equilibrium.

\section{Entropy production}

If the probability $\rho$ is found as a function of time
by solving the FPK equation, then $S$ is obtained as a function of time.
Deriving (\ref{9}) with respect to time, we find
\beq
\frac{dS}{dt} = -k \int \frac{\partial\rho}{\partial t} \ln\rho\, dxdp.
\label{10x}
\eeq
There is another part involving the time derivative of $\ln\rho$
but it vanishes if we take into account that $\rho$ is normalized
at any time, a result given by equation (\ref{3}).

Replacing the time derivative of $\rho$ in (\ref{10x}), by using the 
FPK equation in the form (\ref{7}), we find
\beq
\frac{dS}{dt} =  k \int \frac{\partial J}{\partial p} \ln\rho\, dxdp.
\label{10}
\eeq
Again the part involving the Poisson brackets vanishes
by an integration by parts. 
The entropy is not a conserved quantity like the energy
and as a consequence its variation with time is not equal
to the flux of entropy. In other terms, the right-hand side
of (\ref{1}) cannot be identified as the flux of entropy. 
In addition to the flux of entropy, there is another
contribution related to the creation of entropy.
This contribution is the rate of how entropy is being generated
or created, which is called the rate of entropy production, denoted
by $\Pi$. Thus we variation of the entropy of a system with 
time is written as 
\beq
\frac{dS}{dt} = \Pi - \Phi,
\label{13}
\eeq
where $\Phi$ is the flux of entropy {\it from} the system {\it to}
the outside.
The next step is to define or postulate the expression for
one of the two quantities, $\Pi$ or $\Phi$. Once one of them
is given, the other is obtained by observing that their difference
should be equal to the right-hand side of (\ref{10}).

The rate of entropy production $\Pi$ is a quantity that vanishes when the
thermodynamic equilibrium sets in and gives a measure of the
deviation from equilibrium. As we have seen above, the vanishing
of the irreversible probability current $J$ is a condition for
the the thermodynamic equilibrium. Since the entropy production
is a nonnegative quantity and vanishes when $J$ vanishes,
it should be related to $J^2$. The expression for the $\Pi$
that we are about to introduce meets this two conditions. 

Let $\rho_0$ be the probability distribution that makes 
the irreversible probability current $J$ to vanish. Writing $J$
in the form
\beq
\frac{J}{\rho} = D - \frac{\Gamma}2 \frac{\partial\ln\rho}{\partial p},
\label{45}
\eeq
this condition is equivalent to
\beq
D = \frac{\Gamma}2\frac{\partial \ln \rho_0}{\partial p}.
\label{46}
\eeq
The probability $\rho_0$ does not need to be necessarily the
equilibrium probability distribution $\rho_e$ because we are not
requiring that $\{H,\rho_0\}$ vanishes as it occurs with $\rho_e$.
In analogy with the right-hand side of equation (\ref{10}),
we define the rate of entropy production by
\beq
\Pi =  k \int \frac{\partial J}{\partial p}
(\ln\rho - \ln\rho_0) dxdp.
\label{47}
\eeq
If we integrate by parts and use the relation
\beq
\frac{J}{\rho} = \frac{\Gamma}2\frac{\partial}{\partial p}
(\ln \rho_0 - \ln \rho),
\eeq
that follows from (\ref{45}) and (\ref{46}), we reach the expression
\beq
\Pi = \frac{2k}{\Gamma} \int \frac{J^2}{\rho} dxdp.
\label{47a}
\eeq
We see that the rate of entropy production is the integral
of an expression proportional to $J^2$, as desired.
It is nonnegative and vanishes in equilibrium, when $J=0$, that is,
\beq
\Pi \geq 0,
\eeq
which is a brief statement of the second law of thermodynamics.

The flux of entropy $\Phi$ is obtained from $dS/dt=\Pi-\Phi$
by using the expressions (\ref{10}) and (\ref{47}),
\beq
\Phi = - k \int \frac{\partial J}{\partial p}\ln\rho_0 \,dxdp.
\eeq
Performing an integration by parts and using (\ref{46}), we find
\beq
\Phi = \frac{2k}\Gamma \int J D dxdp.
\label{48}
\eeq

The flux of entropy can also be written as
\beq
\Phi = \frac{2k}{\Gamma}\la D^2\ra + k \la \frac{\partial D}{\partial p}\ra,
\label{20}
\eeq
after the replacement of $J$, given by (\ref{7}) in (\ref{48})
and performing an integration by parts in the second term.
This is an interesting form for the flux of entropy because it
can be understood as an average over the probability distribution
$\rho$, which is not the case of the rate of entropy production.

From the expressions for the flux of entropy $\Phi$ and the rate of
entropy production $\Pi$, we draw an important conclusion concerning
the Liouville equation (\ref{12}). Since this equation can be understood
a the FPK equation without the irreversible probability current $J$,
and since $\Phi$ and $\Pi$ vanishes if $J=0$, it follows that
the Liouville equation predicts no entropy production nor
flux of entropy, and the entropy $S$ is constant in time.
If the Liouville equation is employed to describe a closed
system that approaches equilibrium, but initially is out
or equilibrium, then this prediction of the Liouville equation
is in contradiction with thermodynamics which predicts
an increase of entropy with time.

Let us consider in the following that $D$ and $\Gamma$ are
related by (\ref{5}) where $\rho_e$ is the canonical Gibbs
distribution (\ref{1}) so that the FPK equation describes
a system in contact with a thermal reservoir and at equilibrium.
Replacing the results (\ref{12a}) and (\ref{12b}) in (\ref{6}),
the expression for the irreversible probability current becomes
\beq
J =  -\gamma \left(p \rho + m k T \frac{\partial\rho}{\partial p}\right).
\label{6a}
\eeq
Replacing the expression for $D$ given by (\ref{5}) in the
equation (\ref{48}), we find 
\beq
\Phi = k \int J\frac{\partial\ln\rho_e}{\partial p}  dxdp
= - \frac{1}{T} \int J\frac{\partial H}{\partial p}  dxdp.
\label{10e}
\eeq
Comparing with (\ref{8a}), we see that the entropy flux
and the heat flux are related by
\beq
\Phi = -\frac1T \Phi_q.
\label{73}
\eeq
Using equations (\ref{8b}) and (\ref{13}) we get the following
relation 
\beq
\frac{dU}{dt} - T\frac{dS}{dt} = - T\Pi,
\label{29}
\eeq 
valid at any time. Near equilibrium, $\Pi$ vanishes faster then
the other two time derivatives and 
\beq
\frac{dU}{dt} = T \frac{dS}{dt},
\label{30}
\eeq 
or $dU=TdS$ which is the Clausius relation, valid at thermodynamic
equilibrium. We remark that $T$ here is the temperature of the
heat reservoir. The temperature of the system is
$(\partial U/\partial S)=T^*$ if $U$ could be written as a function of $S$. 
In out of equilibrium, when $\Pi\neq0$, this is not possible,
but in equilibrium, in view of the relation $dU=TdS$, then $U$ becomes
a function of $S$. The relation $dU=TdS$ is translated into
$T=(\partial U/\partial S)$, which implies $T=T^*$, and $T$ becomes also
the temperature of the system.

It is worth mentioning that the variation in time of the
free energy $F$, defined by $F=U-TS$, at $T$ constant, is
\beq
\frac{dF}{dt}=-T\Pi.
\eeq 
which follows form (\ref{29}). Since $\Pi\geq 0$,
then $dF/dt\leq0$ and $F$ decreases monotonically towards
its equilibrium value. This inequality can also be viewed as 
the H theorem of Boltzmann. Defining the Boltzmann $H$ by
\beq
H = \int \rho\ln\frac{\rho}{\rho_e} dxdp,
\eeq
we see that it is equal to $-\beta F$ plus a constant. Then,
it follows from $dF/dt\leq0$ that $dH/dt\geq0$, which is
the H theorem of Boltzmann.

\section{Work}
\label{work}

The specific systems that we have consider so far are those that 
exchange only heat with the environment and are described by
the FPK equation (\ref{7}).
Now we wish to consider the
case where the systems are also subject to external forces.
The appropriate way to treat this case is to add to the
ordinary force appearing in the FPK equation (\ref{2})
an external force $F^e$, so that $f$ now reads
\beq
f = F^c + F^e + D.
\eeq 
Repeating the reasoning leading to (\ref{7}) from (\ref{2}),
we reach the evolution equation
\beq
\frac{\partial\rho}{\partial t} = \{H,\rho\}
- \frac{\partial F^e\rho}{\partial p}
- \frac{\partial J}{\partial p},
\label{7a}
\eeq
where $J$ is the irreversible probability current, given by (\ref{6a}),
which is the one appropriate for the contact with a heat reservoir
at a temperature $T$.

Due to the presence of the external force, 
the variation in time of the energy has another
contribution in addition to the flux of heat,
\beq
\frac{dU}{dt} = \Phi_q - \Phi_w,
\label{41}
\eeq
where $\Phi_q$ is the heat flux into the system and $\Phi_w$
is the work performed by the system per unit time
against the the external forces, or power.

To determine the variation of energy with time, 
we proceed in the same way as we did to derive (\ref{8b})
from the evolution equation (\ref{2}), but now we use the
evolution equation (\ref{7a}). The result is 
the equation (\ref{41}) where $\Phi_q$
is the expression (\ref{8a}) and the power $\Phi_w$ is
\beq
\Phi_w = -\int F^e\rho\frac{\partial H}{\partial p} dxdp.
\label{57}
\eeq

The equation (\ref{10}) for the variation of entropy with time
remains unchanged by the addition of the external force.
To see this we replace the expression (\ref{7a}) into (\ref{10x}).
The term involving the Poisson brackets vanishes as we
have already seen. The term involving the external force is
\beq
k \int \frac{\partial F^e\rho}{\partial p}\ln \rho dx dp
= - k \int  F^e \frac{\partial \rho}{\partial p} dx dp,
\eeq
where we have performed an integration by parts.
But this integral also vanishes if we assume that
$F^e$ does not depend on $p$.

The rate of entropy production 
is defined by (\ref{47a}), and considering that the expression
for $dS/dT$ remains unchanged, as we have just seen, so does
the expression (\ref{20}) for the entropy flux.
As these relations are not modified by the presence of
the external force, then the relation
$\Phi=-\Phi_q/T$, as expressed by equation (\ref{73}),
between the entropy flux and the heat flux, valid for a 
system is in contact we a heat reservoir at a temperature $T$,
also remains unchanged.

Taking into account that $dS/dt=\Pi-\Phi$ and that $dU/dt=\Phi_q-\Phi_w$,
we reach the following relation
\beq
\frac{dU}{dt} - T \frac{dS}{dt} = - \Phi_w - T\Pi.
\eeq
Considering a process in which $T$ is constant, then the left
hand side is $dF/dt$ where $F=U-TS$ is the free energy, that is,
\beq
\frac{dF}{dt} = - \Phi_w - T\Pi.
\eeq
Integrating in time, between $t_1$ and $t_2$, we find 
\beq
\Delta F = - W - T\int_{t_1}^{t_2} \Pi dt.
\label{39}
\eeq
where $W$ is the work performed by the external force,
\beq
W = \int_{t_1}^{t_2} \Phi_w \,dt,
\label{87}
\eeq
which is the time integration of the power $\Phi_w$.
Since $\Pi\ge0$, it follows that
\beq
\Delta F \leq - W,
\label{40}
\eeq
that is, the variation of the free energy is smaller than
the work done {\it on} the system. The equality holds in an equilibrium
process, when $\Pi=0$.

The following remark is in order. When the system is in contact
with a heat reservoir at a certain temperature $T$, it does not mean
that $T$ is the temperature of the system, as we have pointed out
in the remark just below equation (\ref{30}). 
If the rate of entropy production is nonzero, 
no temperature could be assigned to the system and $F=U-TS$
could not be,  strictly speaking, the free energy of the system
although $U$ and $S$ are the energy and entropy of the system.
But we may suppose that for $t\leq t_1$ and $t\geq t_2$, the system
is in equilibrium, in which case $\Pi$ is nonzero only for $t_1<t<t_2$.
Within this scenario, $T$ can be considered the temperature
of the system at time $t_1$ and at time $t_2$, 
and $F$ at these two instants of time
will be the free energy of the system, and the relation $\Delta F$
in (\ref{39}) will represent the difference in the free energies of the
system.

The derivations that we have just carried out, 
such as that of the inequality (\ref{40}), made no restriction
on the type of external force. It could be a nonconserved force
or a time dependent force. This latter type of external force 
happens, for instance, when the system is driven at our will.

\section{Harmonic oscillator}

Let us apply the results we have found so far to the case
of a harmonic oscillator for which $F^c=-K x$ and
\beq
H = \frac{p^2}{2m} + \frac12 K x^2.
\label{58}
\eeq
It is in contact with a heat reservoir at a temperature
$T$ so that the FPK equation is 
\beq
\frac{\partial\rho}{\partial t} =
- \frac{p}{m}\frac{\partial\rho}{\partial x}
+ K x \frac{\partial\rho}{\partial p}
+ \gamma \frac{\partial p\rho}{\partial p}  
+ \frac{m \gamma}\beta \frac{\partial^2\rho}{\partial p^2}.
\label{2b}
\eeq

The FPK equation can be solved exactly by assuming the
following Gaussian form for the probability distribution
\beq
\rho = \frac1Z e^{-(ax^2+bp^2+2cxp)/2},
\label{21}
\eeq
where the parameters $a$, $b$, $c$ depend on time and
\beq
Z = \frac{2\pi}{\sqrt{ab-c^2}}.
\eeq
The solution is given in the appendix \ref{B}, where we find
the parameters $a$, $b$, and $c$ as functions of time. 

From the probability distribution (\ref{21}),
we determine the covariances,
\beq
\la x^2\ra = \frac{b}{ab-c^2},
\eeq
\beq
\la p^2\ra = \frac{a}{ab-c^2},
\eeq
\beq
\la xp\ra =  \frac{-c}{ab-c^2},
\eeq
and other properties.
The energy is
\beq
U = \frac1{2m}\la p^2\ra + \frac{K}2 \la x^2\ra,
\eeq
and the entropy is found by using its definition and is
\beq
S = k + \ln2\pi - \frac12 \ln(ab-c^2).
\eeq

To determine $dS/dt$, $\Pi$ and $\Phi$, we need to find $J$,
which is defined by (\ref{6a}). In the present case it reads
\beq
J =  \frac{m\gamma}{\beta}(cx + bp - \frac{\beta}{m}p)\rho.
\eeq
From $J$ we find 
\beq
\frac{dS}{dt} =   \frac{k m\gamma}{\beta} b - k\gamma,
\eeq
\beq
\Phi = \frac{k\beta\gamma}{m}\la p^2\ra - k \gamma,
\eeq
\beq
\Pi = \frac{k\beta\gamma}{m}\la p^2\ra +  \frac{km\gamma}\beta b
- 2k\gamma,
\eeq
and it becomes clear that $dS/dt=\Pi-\Phi$.

From the asymptotic values of the parameter $a$, $b$, and $c$,
given in the appendix \ref{B}, which are $a=K\beta$, $b=\beta/m$, and
$c=0$, we find the equilibrium values of the various quantities
which are $\la xp \ra = 0$,
\beq
\frac1{2m} \la p^2\ra
=\frac12 K\la x^2\ra = \frac1{2\beta} = \frac12 kT,
\eeq
\beq
U = \frac1\beta = kT,
\eeq
\beq
S = k + \ln 2\pi - \frac12 \ln \frac{K\beta^2}{m},
\eeq
$dS/dt=0$, $\Phi=0$, and $\Pi=0$.
As the parameters $a$, $b$, and $c$ decay exponentially to
their asymptotic values, so do the properties obtained above. 

We remark that the probability distribution approaches the
equilibrium probability distribution
\beq
\rho_e = \frac1Z e^{-\beta H},
\eeq
where $H$ is the energy function (\ref{58}), and the system
thermalizes properly.

\section{Quantum evolution equation}

To extend the stochastic approach developed above
to the quantum case we need to provide a quantum
version of the evolution equation. One way of setting 
up the quantum version is to use a procedure
known as canonical quantization, which amounts to
replace the Poisson brackets of classical mechanics by
a quantum commutator. For the case of just one degree of freedom
that we are considering here, the Poisson brackets 
between $A$ and $B$ are given by
\beq
\{A,B\} = \frac{\partial A}{\partial x}\frac{\partial B}{\partial p}
- \frac{\partial B}{\partial x}\frac{\partial A}{\partial p}.
\eeq
The canonical quantization is obtained by performing
the replacement
\beq
\{A,B\} \to \frac1{i\hbar} [\hat{A},\hat{B}],
\eeq
where $\hbar$ is the Planck constant and the quantities
$\hat{A}$ and $\hat{B}$ on the right are understood as quantum operators,
and $[\hat{A},\hat{B}]=\hat{A}\hat{B}-\hat{B}\hat{A}$.

From the quantization rule above, we obtain two useful rules.
The first is obtained by setting $A=x$ in the Poisson brackets,
which gives
\beq
\{x,B\} = \frac{\partial B}{\partial p}.
\eeq
Using the quantization rule, we obtain
\beq
\frac{\partial B}{\partial p} \to \frac1{i\hbar} [\hat{x}, \hat{B}].
\label{59a}
\eeq
In an analogous way, if we set $B=p$, we find
\beq
\{A,p\} = \frac{\partial A}{\partial x},
\eeq
and using the quantization rule, we obtain
\beq
\frac{\partial A}{\partial x} \to \frac1{i\hbar} [\hat{A}, \hat{p}].
\label{59b}
\eeq
We remark that $\hat{x}$ and $\hat{p}$ on the right-hand sides
of (\ref{59a}) and (\ref{59b}) are 
quantum operators representing the position and the momentum
of a particle, and we recall that according to quantum
mechanics $[\hat{x},\hat{p}]=i\hbar$.

The last two rules are useful in the transformation of 
a differential equation such as the FPK equation
into a quantum equation. It should be remarked however
that the equation obtained by this procedure is not
a mathematical derivation of the quantum equation from
the classical equation. In fact, the opposite is true.
From the quantum equation one reaches the classical
equation by taking the classical limit. Thus, the quantization
rules should be used as a guide to find a quantum equation
which at the end should be introduced as a postulate.

It is usual to use the hat symbol to denote a quantum operator
as we have done above. But from now on we will drop the hat symbol and
denote an operator by a letter without the hat. Thus the
position and momentum operator, for instance, will be denoted by
$x$ and $p$.

Let us consider the FPK equation in the form (\ref{7}).
According to the quantization rules, the quantum evolution
equation is 
\beq
\frac{\partial\rho}{\partial t} = \frac1{i\hbar}[H,\rho]
- \frac1{i\hbar} [x,J],
\label{22}
\eeq
where now, $\rho$, $H$, and $J$ are quantum operators.
Since the quantum operators can be represented by
matrices most properties of quantum operators are better
understood if stated in terms of matrices. 
For instance the density operator $\rho$, which is the quantum
version of the probability density distribution, 
holds the following property:
its diagonal elements are nonnegative and the sum of the
diagonal elements, its trace, equals the unity,
\beq
{\rm Tr}\rho = 1.
\label{23}
\eeq
The operator $H$ is the quantum energy function, or the
Hamiltonian, given by
\beq
H = \frac1m p^2 + V,
\eeq
where $V$ is a function of $x$. We remark that without $J$
the equation reduces to the quantum Liouville equation
of quantum statistical mechanics.

The property (\ref{23}) is the analogue of the normalization
of the probability density that we used before and thus it
should be preserved in time. To see that this is the case
let us take the trace of the right-hand side of equation (\ref{22}).
There are two terms to be considered and both vanish 
because each one is the trace of a commutator, and the
trace of a commutator vanishes. Therefore,
the left hand side, which is the time derivative of the 
trace of $\rho$, vanishes and $\rho$ should be constant in time.

The trace of a commutator vanishes because ${\rm Tr}(AB)={\rm Tr}(BA)$.
This is a particular case of the cyclic property of the trace
${\rm Tr}(ABC)={\rm Tr}(BCA)$. This cyclic property allows us to
write the following property
\beq
{\rm Tr}([A,B]C) = {\rm Tr}(A[B,C]),
\label{24}
\eeq
that we will employ further on.

\section{Energy and entropy}

The average $U=\la H\ra$ of the energy function $H$ with respect to the 
density operator $\rho$ is given by
\beq
U = {\rm Tr}(H\rho),
\eeq
and is the quantum analog of the integral in (\ref{17}).
Deriving it with respect to time, 
\beq
\frac{dU}{dt} = {\rm Tr} \left(H \frac{\partial \rho}{\partial t}\right),
\eeq
and using the evolution equation (\ref{22}), we find
\beq
\frac{dU}{dt} = - \frac1{i\hbar} {\rm Tr}(H [x,J]),
\eeq
where the term involving the commutator $[H,\rho]$ vanishes
owing to the property (\ref{24}). Using again this same property
we get
\beq
\frac{dU}{dt} = \frac1{i\hbar} {\rm Tr}([x,H]J).
\eeq
The right-hand side is interpreted as the heat flux into the system
\beq
\Phi_q = \frac1{i\hbar} {\rm Tr}([x,H]J),
\label{27}
\eeq
and
\beq
\frac{dU}{dt} = \Phi_q.
\eeq

The definition of entropy for the quantum case is that introduced by
von Neumman, 
\beq
S = -k {\rm Tr}(\rho \ln\rho),
\eeq
and corresponds to the extension of the Gibbs entropy to 
the quantum case. Deriving this equation with respect to time
we find
\beq
\frac{dS}{dt} = -k {\rm Tr}\left(\frac{\partial\rho}{\partial t}
\ln\rho\right),
\eeq
where the term involving the derivative of $\ln\rho$
vanishes in view of the normalization property (\ref{23}).
Using the evolution equation, we find
\beq
\frac{dS}{dt} = \frac{k}{i\hbar}{\rm Tr}([x,J]\ln\rho),
\label{25}
\eeq
where again the term involving the commutator $[H,\rho]$ vanishes
owing to the property (\ref{24}).

\section{Rate of entropy production}

The right-hand side of the equation (\ref{25}) is not equal to
to the flux of entropy because the entropy is not a conserved quantity.
There is another source of entropy which comes from dissipation
inside the system. Thus as before the time variation of the
entropy has two term,
\beq
\frac{dS}{dt} = \Pi - \Phi,
\label{26}
\eeq
where $\Pi$ is the rate of entropy production and $\Phi$
is the flux of entropy {\it from} the system to the outside.

Guided by the classical version, the production of entropy is
defined as follows. The quantum irreversible current $J$ has not
been defined yet but it is expressed in terms of the density
operator, that is, $J(\rho)$. Let us denote by $\rho_0$ the
quantity such that $J(\rho_0)=0$. If the commutator of $\rho_0$
with $H$ also vanishes then $\rho_0$ is identified as the density at
thermodynamic equilibrium. However, here we do not demand that
this commutator vanishes. The rate of entropy production is defined as
\beq
\Pi = \frac{k}{i\hbar}{\rm Tr}\{[x,J](\ln\rho -\ln\rho_0\},
\eeq
and it becomes clear that $\Pi$ vanishes whenever $J$ vanishes.
Taking into account (\ref{25}) and (\ref{26}), the expression for
the flux of entropy $\Phi$ is 
\beq
\Phi = - \frac{k}{i\hbar}{\rm Tr}\{[x,J]\ln\rho_0\}.
\eeq

Let us assume that the irreversible current $J$,
which has not yet been specified,
is so defined that the evolution equation
describes a system in contact with a heat reservoir at temperature
$T$. In thermodynamic equilibrium $J$ vanishes and $\rho_0$ 
should be identified as corresponding to the Gibbs canonical
distribution which in the quantum case reads
\beq
\rho_0 = \frac1Z e^{-\beta H},
\label{34}
\eeq
where
\beq
Z = {\rm Tr} (e^{-\beta H}).
\eeq
Replacing $\rho_0$ in the expressions for $\Pi$ and $\Phi$, we find
\beq
\Pi = \frac{k}{i\hbar}{\rm Tr}\{[x,J](\ln\rho + \beta H\},
\eeq
\beq
\Phi = \frac{k\beta}{i\hbar}{\rm Tr}([x,J] H).
\label{28}
\eeq

Now let us compare equations (\ref{28}) and (\ref{27}).
We see that $\Phi$ and $\Phi_q$ are related by
\beq
\Phi = - \frac1T \Phi_q,
\eeq
which is the thermodynamic relation that should exist between 
the flux of entropy $\Phi$ and the heat flux $\Phi_q$ when a
system is in contact with a heat reservoir at a temperature $T$.
Other thermodynamic relations that we have obtained for
the classical case, such those given by equations (\ref{29}) and (\ref{30}),
can also be shown to be valid in the quantum case.

\section{Irreversible current}

The quantum irreversible current $J$ has not yet been specified.
Here we will choose it by applying the rules of the canonical
quantization to the expression (\ref{6}). The form chosen for
the irreversible current is
\beq
J =  \frac12(D^\dagger\rho + \rho D)
- \frac{\Gamma}2 \frac1{i\hbar}[x,\rho],
\label{31}
\eeq
where $D$ is a quantum operator representing the dissipative force
and $\Gamma$ is a real constant as in the classical case.
Notice that we have written a symmetrized form for the product of
the dissipative force and the density operator. In one of the
products, we have used the Hermitian conjugate of $D$, denoted
$D^\dagger$, so that the whole expression is Hermitian.

The matrix $A^\dagger$ that represents the Hermitian
conjugate of an operator is obtained from the matrix $A$ that
represent the operator by transposing the elements of the matrix
and taking the complex conjugate of each element. If $A_{ij}$ and
$(A^\dagger)_{ij}$ represent an element of these two matrices then
$(A^\dagger)_{ij}=(A_{ji})^*$. A Hermitian operator is the operator
which is equal to its Hermitian conjugate. An important property
of such an operator is that its eigenvalue are real.

If we wish to describe a system that at long times reaches the
thermodynamic equilibrium then $D$ and $\Gamma$ should have
a relation such that $J$ vanishes when $\rho$ is the equilibrium
density operator $\rho_e$. Imposing the vanish of $J$ when
$\rho$ is replaced by $\rho_e$ we find the condition
\beq
\frac12(D^\dagger\rho_e + \rho_e D)
= \frac{\Gamma}2 \frac1{i\hbar}[x,\rho_e].
\eeq
The solution for $D$ is
\beq
D = \frac{\Gamma}2 \frac1{i\hbar}\rho_e^{-1}[x,\rho_e]
= \frac{\Gamma}2 \frac1{i\hbar}(\rho_e^{-1}x\rho_e - x),
\label{35}
\eeq
which is understood as the relation between
dissipation, described by $D$, and noise or fluctuations,
described by $\Gamma$.
In equilibrium there must be a relation between
dissipation and fluctuations.

That (\ref{35}) is a solution can be verified by substitution and
using the property
that the Hermitian conjugate of a product of two operators equals
the product of the conjugate of each operator in the inverse order.
In the present case, $(\rho_e D)^\dagger=D^\dagger \rho_e$ because
$\rho_e$ is Hermitian.

Next we seek for $J$ that could describe the contact of the system
with a heat reservoir. In this case
\beq
\rho_e = \frac1Z e^{-\beta H},
\eeq
which replace in (\ref{35}) gives
\beq
D = \frac{\Gamma}2 \frac1{i\hbar}(e^{\beta H}xe^{-\beta H} - x).
\label{35a}
\eeq
This form of dissipation is certainly not the form of the classical
dissipation found above which is proportional to the momentum.
However, at high temperatures this is the
case. If we expand the terms between parentheses in the right-hand
side of (\ref{35a}) up to terms of order $\beta$, we find
$D = - \gamma p$ where $\gamma=\Gamma\beta/2m$, or
\beq
\Gamma = \frac{2\gamma m}{\beta}.
\label{36}
\eeq
For an arbitrary temperature, the dissipation, according to
the present approach, is not proportional to the momentum
and is given by (\ref{35a}), which we write, by using (\ref{36}), as
\beq
D = - \gamma g,
\eeq
where 
\beq
g = - \frac{m}{i\hbar\beta}(e^{\beta H}xe^{-\beta H} - x).
\label{35b}
\eeq

Replacing the irreversible current (\ref{31}) into the evolution
equation (\ref{22}) we may write it in the more explicit form
\beq
\frac{\partial\rho}{\partial t} = \frac1{i\hbar}[H,\rho]
+ \frac{\gamma }{2i\hbar} [x,g^\dagger\rho + \rho g]
- \frac{\gamma m}{\beta\hbar^2} [x,[x,\rho]],
\label{37}
\eeq
which we call the quantum FPK equation.

\section{Quantum harmonic oscillator}

For a quantum Harmonic oscillator the energy function is
\beq
H = \frac1{2m} p^2 + \frac12 m\omega^2 x^2,
\label{63}
\eeq
where $\omega$ is the frequency of oscillations.
To find the solution of the quantum FPK equation from which we may
determine the thermodynamic properties it is necessary to know
the explicit expression of the dissipation force $D=-\gamma g$,
that is, we need to know $g$ as a function of $x$ and $p$.
For the harmonic oscillator we show in the appendix \ref{C} that
\beq
g = a p + i b x,
\eeq
where $a$ and $b$ are real numbers given by
\beq
a = \frac{1}{\beta\hbar\omega}\sinh\beta\hbar\omega,
\qquad
b = \frac{m}{\beta\hbar}(\cosh\beta\hbar\omega - 1).
\eeq

A solution of the quantum FPK equation (\ref{37})
can be obtained by a method similar to that used in the
classical case, which is to assume a solution of the form
\beq
\rho = \frac1Z e^{-(ax^2+bp^2+cxp+cpx)/2},
\eeq
where $a$, $b$, and $c$ are real constant that depends on time.
Here we will limit ourselves to write down the equations for
the covariances and determine their asymptotic values, which
are the values at thermodynamic equilibrium. 
The time dependent solution can be found in the reference
\cite{oliveira2016}.
Multiplying the quantum FPK equation successively by $x^2$, $p^2$, 
and $xp$, and taking the trace, we obtain the following equations
for the covariances
\beq
\frac{d}{dt}\la p^2\ra = 
-m\omega^2 (\la px\ra + \la xp\ra) + \hbar b\gamma
-2a\gamma \la p^2\ra + \frac{2\gamma m}{\beta},
\label{23a}
\eeq
\beq
\frac{d}{dt}\la x^2\ra = \frac{1}{m} (\la px\ra + \la xp\ra),
\label{23b}
\eeq
\beq
\frac{d}{dt}\la xp\ra = \frac{1}{m} \la p^2\ra
-m\omega^2\la x^2\ra
-\frac{a\gamma}2(\la px\ra + \la xp\ra).
\label{23c}
\eeq
The equation for $\la px\ra$ is not needed because
$\la xp\ra-\la px\ra=i\hbar$.

At the stationary state, which is a state of thermodynamic
equilibrium we find
\beq
\la xp\ra =-\la xp\ra= \frac{i\hbar}{2},
\eeq
\beq
\frac1{2m}\la p^2\ra = \frac12m\omega^2\la x^2\ra
= \frac12\hbar\omega\left(\frac{1}{e^{\beta\hbar\omega}-1}+\frac12\right).
\eeq
From these results one reaches the expected expression for 
the average energy of a quantum oscillator,
\beq
\la H\ra
= \hbar\omega\left(\frac{1}{e^{\beta\hbar\omega}-1}+\frac12\right).
\eeq

We remark that the probability distribution approaches the
equilibrium probability distribution
\beq
\rho_e = \frac1Z e^{-\beta H},
\eeq
where $H$ is the quantum energy function (\ref{63}), and the system
thermalizes properly.

\section{Multiple degrees of freedom}

Up to this point, we have considered systems
with just one degree of freedom. Here we wish to consider
the case of a system with multiple degrees of freedom. 
The derivations of the results for the present case parallel
those obtained for one degree of freedom and will not be
shown in full detail. We restrict ourselves to the classical case 
but the quantum case can be obtained in a way
similar to the case of one degree of freedom
and can be found in reference \cite{oliveira2016}.

An appropriate treatment of a system with many degrees of freedom
begins with the generalization of the FKP equation (\ref{7a})
for this case. As we have seen, this equation
describes a system in contact with a heat reservoir 
and is subject to an external force.
The generalization that we give below
is the one appropriate to describe
a system in contact with multiple reservoirs at
distinct temperatures and subject to several forces.
We denote by $x_i$ a Cartesian 
component of the position and by $p_i$ the
respective component of the momentum related to a certain
degree of freedom. The energy function
is a sum of kinetic energy and a potential energy,
\beq
H = \sum_i \frac{p_i^2}{2m} + V.
\eeq
The FPK equation, which governs the time evolution of
the probability density $\rho$, now reads
\beq
\frac{\partial\rho}{\partial t} = \{H,\rho\}
- \sum_i\frac{\partial F_i^e\rho}{\partial p_i}
- \sum_i\frac{\partial J_i}{\partial p_i},
\label{76}
\eeq
where $F_i^e$ are the Cartesian components of the external force,
which may be nonconservative and time dependent, and
\beq
J_i = D_i \rho - \frac{\Gamma_i}2 \frac{\partial \rho}{\partial p_i}
\label{64}
\eeq
are the components of the dissipative probability current.
We choose the dissipation force $D_i$ to be of the
usual form $D_i=-\gamma p_i$ and $\Gamma_i=2mkT_i$ so 
that we may interpret the FPK equation of describing
a system in contact with several heat reservoirs at
temperatures $T_i$. Therefore,  
\beq
J_i =  -\gamma \left(p_i \rho
+ m k T_i \frac{\partial\rho}{\partial p_i}\right)
\eeq

In the absence of external forces and if all heat baths 
have the same temperature $T_i=T$, then the stationary
state is a state of thermodynamic equilibrium because in
this case $J_i$ vanishes for all $i$. Indeed, if we replace
the Gibbs probability distribution 
\beq
\rho_e = \frac1Z e^{-\beta H}
\eeq
where $\beta=1/kT$, in the expression for $J_i$ we see that it
vanishes. We remark that the Poisson brackets also vanish.

The time variation of the energy $U=\la H\ra$ has the
same form of equation (\ref{41}),  
\beq
\frac{dU}{dt} = \Phi_q - \Phi_w,
\label{77}
\eeq
but now $\Phi_q$ is a sum of the heat fluxes coming from each reservoir,
that is,
\beq
\Phi_q = \sum_i \Phi_{qi},
\eeq
where each heat flux has the form (\ref{8a}), with
$J$ replaced by $J_i$, 
\beq
\Phi_{qi} = \int J_i\frac{\partial H}{\partial p_i} dxdp.
\label{8x}
\eeq
Using the relation $\partial H/\partial p_i=p_i/m$
and performing an integration by parts, the heat flux 
can be written as
\beq
\Phi_{qi} = -\gamma \left(\frac1m \la p_i^2\ra -  k T_i\right).
\label{68}
\eeq
It becomes clear that if $kT_i/2$ is larger than
the average kinetic energy $\la p_i^2\ra/2m$ then $\Phi_{qi}>0$, 
and heat flows {\it into} the system, otherwise, heat flows
{\it from} the system to the heat reservoir.

The expression for the power $\Phi_w$ is similar to that
of equation (\ref{57}) but now there is a sum over all 
components of the force
\beq
\Phi_w = -\sum_i \int F_i^e\rho\frac{\partial H}{\partial p_i} dxdp.
\label{8w}
\eeq
Using again the relation $\partial H/\partial p_i=p_i/m$
and performing an integration by parts, the power
can be written as
\beq
\Phi_w = - \frac1m\sum_i \la F_i^e p_i\ra.
\label{67}
\eeq
It is useful to understand that the forces $F_i^e$
are being acted by an external agent which is
a power device. The role played by the power device in
relation to the transfer of
mechanical work is analogous to the role played by the
heat reservoir in relation to the transfer of heat.
We see from (\ref{67}) that the power has the usual form of
a force multiplied by the velocity. If $F_i^e$ and $p_i$
have the same sign, work is performed by the power device
onto the system, otherwise, the system performs work on the
power device. 

The heat flux and the power in (\ref{77}) might be understood 
as functions of the time so that we may write
$\Phi_q=dQ/dt$ and $\Phi_w=dW/dt$ in which case equation
(\ref{77}) reduces to the form
\beq
dU = dQ - dW,
\eeq
which is the usual way of writing the conservation of
energy. However, it should be remarked that both
$dQ$ and $dW$ are not in general exact differential,
although $dU$ is.
The concepts of exact and inexact differentials are
commented in the appendix \ref{D}.

The variation of entropy with time is
\beq
\frac{dS}{dt} = \Pi - \Phi,
\eeq
where the rate of entropy production $\Pi$ and the entropy
flux $\Phi$ are generalization
of the equations (\ref{47a}) and (\ref{20}) for the present case, 
\beq
\Pi = \frac1{m\gamma}\sum_i \frac1{T_i} \int \frac{J_i^2}{\rho} d\xi,
\eeq
where the integration extends over the space of states, and
\beq
\Phi = \sum_i \frac{\gamma}{T_i}\left(\frac{1}{m}\la p_i^2\ra - kT_i\right).
\eeq
In view of the equation (\ref{68}), we see that the flux of
entropy is related to the heat fluxes by
\beq
\Phi = - \sum_i \frac{\Phi_{qi}}{T_i}.
\label{69}
\eeq
Using this relation, we may derive again
all the results involving the free energy obtained
in section \ref{work}, including the inequality \ref{40}.

\section{Nonequilibrium steady state}

Here we wish to show by an example 
that the present approach is indeed
capable of describing systems displaying nonequilibrium
steady states. That is, for long times the systems reaches
a stationary state in which the irreversible currents are
nonzero and entropy is permanently being created.
One way of setting a system in a nonequilibrium steady
state is to place the system in contact with
heat reservoirs with different temperatures. Another
possibility is to place the system under the action
of a power device. The two possibilities are embodied
in the development made in the previous section. 

We examine a system with two degrees of freedom.
The potential energy is harmonic and the energy function is
\beq
H = \frac1{2m}(p_1^2+p_2^2) + \frac12 K(x_1^2+x_2^2) - L x_1x_2,
\label{83}
\eeq
so that the conservative forces
are linear, $F_1=-Kx_1 + Lx_2 $ and $F_2=-Kx_2+Lx_1$.
In addition to the conservative forces, the system is 
under the action of a nonconservative external force
given by
\beq
F_1^e= - cx_2, \qquad\qquad F_2^e= cx_1.
\label{84}
\eeq
According to equation (\ref{67}), 
\beq
\Phi_w = \frac{c}m \la x_2 p_1 \ra - \frac{c}m \la x_1 p_2 \ra,
\label{71}
\eeq
which is minus the power performed by the power device.
Each degree of freedom is understood as being coupled to one
of the two heat reservoirs at the temperatures $T_1$ and $T_2$. 
The fluxes of heat associated to each reservoir are given by
(\ref{68}) and are
\beq
\Phi_{q1} = -\frac{\gamma}m (\la p_1^2\ra -  mk T_1),
\label{70a}
\eeq
\beq
\Phi_{q2} = -\frac{\gamma}m (\la p_2^2\ra -  mk T_2).
\label{70b}
\eeq
From the two heat fluxes we determine the entropy flux
by means of the relation (\ref{69})
\beq
\Phi = - \frac{\Phi_{q1}}{T_1} - \frac{\Phi_{q2}}{T_2}.
\label{70}
\eeq

From now on we wish to determine the above quantities
in the steady state. In the steady state $\Pi=\Phi$
and in view of the relation (\ref{70}), it suffices
to determine the heat fluxes $\Phi_{qi}$ and the
power $\Phi_w$. To find these quantities we need
the covariances $\la x_i p_j\ra$ and $\la p_i^2\ra$
at the steady state. Therefore, we should solve the
FPK equation (\ref{76}), with the energy function $H$
given by (\ref{83}) and external forces given by
(\ref{84}).

In view of the fact that the forces, conservative
and nonconservative, are linear,
it is possible to solve the FPK equation exactly.
The solution is carried out in the appendix \ref{E}
where the covariances at the stationary state are determined.
Replacing the covariances in the expressions (\ref{71}),
(\ref{70a}), and (\ref{70b}), we find
\beq
\Phi_w = -\frac{2c}m {\cal C},
\eeq
\beq
\Phi_{q1} = -\frac{c-L}m {\cal C},
\qquad
\Phi_{q2} = -\frac{c+L}m {\cal C},
\eeq
where 
\beq
{\cal C} = {\cal C}_0(L \Delta T + 2 c T),
\eeq
$T=(T_1+T_2)/2$, $\Delta T = T_1-T_2$, and 
\beq
{\cal C}_0 = \frac{m\gamma k}{2(m\gamma^2 K + L^2 - c^2)}.
\eeq
The range of values of $c$ are such that the 
denominator be positive.

In the steady state the production of entropy equals
the entropy flux which, from (\ref{70}) is given by
\beq
\Pi = \Phi =   \frac{{\cal C}_0}{T_1T_2m} (2c T + L \Delta T)^2
\eeq
which is clearly nonnegative.

Let us analyze the results we have just found
for $L>0$. The various possibilities for the
heat fluxes and power are shown in table \ref{flux}. 
In all cases where $c\neq0$, except one, energy is being dissipated,
that is, work is performed onto the system ($\Phi_w<0$)
which in turn releases it in the form of heat to one or to both heat
reservoirs. The exception is the case in which the system
perform works ($\Phi_w>0$) in which case heat
flows from the hotter reservoir to the colder reservoir
through the system, and the whole system functions as
a heat engine.

\begin{table}
\caption{Heat fluxes and power for the case of the system
defined by the energy function (\ref{83}) and by the external force
(\ref{84}), where $\Delta T=T_1-T_2$. The convention for the fluxes are:
if $\Phi_{q1}$ is positive, heat flows from the reservoir 1
to the system, if $\Phi_{q2}$ is positive, heat flows from the reservoir
2 to the system, if $\Phi_{w}$ is positive, the system performs work
on to the external device. Notice that $\Phi_w=\Phi_{q1}+\Phi_{q2}$.
We are considering
here $L>0$ and $|c|<L$, and that $T_1\geq T_2$.}
\vspace{0.3cm}
\begin{tabular}{|c|c|c|c|c|c|c|}
\hline
$\Delta T$ & $c$ & ${\cal C}$  & $\Phi_w$ & $\Phi_{q1}$ & $\Phi_{q2}$ \\
\hline
0 & + & + & - & + & - \\
0 & - & - & - & - & + \\
+ & 0 & + & 0 & + & - \\
+ & + & + & - & + & - \\
+ & - & + & + & + & - \\
+ & - & - & - & - & + \\
\hline
\end{tabular}
\label{flux}
\end{table}

\section{Path integral}
\label{path}

The approach to the stochastic thermodynamics
that we have developed here is based on
the FPK equation which is understood as an equation
that governs the time evolution of the probability
distribution $\rho$. The solution of the evolution
equation gives $\rho$ at any instant of time. 
For convenience here we denote a state by $\xi$ which
is understood as the collection of positions
and momenta of the particles of the system.

If we solve the FPK equation considering that at the
initial time $t_0$ it was in a certain state $\xi_0$,
then $\rho(\xi,t)$ is understood 
as the conditional probability of finding the system
in state $\xi$ at time $t$, given that it was in the
state $\xi_0$ at time $t_0$, and we denote it by
$P(\xi,t|\xi_0,t_0)$. 

Let us consider now a discretized trajectory, that is,
a trajectory for which
the system is at state $\xi_0$ at an initial time $t_0$, in
$\xi_1$ at time $t_1$, in $\xi_2$ at time $t_2$, ..., and 
in $\xi_n$ at the final time $t_n$. 
The probability of the
occurrence of this discretized trajectory is
a successive product of
\beq
P(\xi_\ell,t_\ell|\xi_{\ell-1},t_{\ell-1}),
\eeq
from $\ell=1$ until $\ell=n$, multiplied by the
probability $P(\xi_0,t_0)$.
Omitting the reference to
the instants of time, the trajectory probability is  
\beq
P(\xi_n|\xi_{n-1})\ldots P(\xi_2|\xi_1) P(\xi_1|\xi_0)P(\xi_0).
\label{81}
\eeq
From now on we consider that the time intervals 
between two successive instants of time are the same
and equal to $\tau$.
It is understood that $\tau$ is small enough so that
the trajectory approaches a continuous trajectory.

Generally speaking the probability of a trajectory 
is a joint probability, which we denote by 
\beq
P(\xi_n,\xi_{n-1},\ldots,\xi_2,\xi_1,\xi_0).
\label{82}
\eeq
The identification of (\ref{82}) with (\ref{81}) defines 
a type of stochastic dynamics associated with the name
of Markov and the approach we are using here,
with the FKP equation as the evolution equation, 
is thus a Markovian stochastic dynamics.

The probability distribution (\ref{82}) is a joint
probability distribution. If we integrate in all variables
except one one them, we find the probability of this
variable. For instance, if we integrate in
$\xi_0,\xi_1,\ldots,\xi_{n-1}$,
we find the probability of $\xi_n=\xi$ at time $t_n=t$,
which is $\rho(\xi,t)$, that is,
\beq
\rho(\xi) = \int ...\int P(\xi,\xi_{n-1},...,\xi_0) d\xi_{n-1}... d\xi_0.
\eeq

In which circumstance should we use the path probability
(\ref{82})? If we wish to find the average $U$ of the energy
function $H(\xi)$ at time $t$, for instance, it suffices to use
the probability distribution $\rho(\xi,t)$. However, if we wish
to find the average of the mechanical work, we should use the
path probability because the work is a path integral.

Let $L$ be the work along a certain trajectory of the
force with components $f_i$,
\beq
L = \sum_i \int_{\rm path} \!\!\! f_i dx_i
\label{86a}
\eeq
where the index 'path' serves to remember that the integral is
an integral along a certain trajectory. 
In a discretized form, the path integral is written the sum
\beq
\int_{\rm path} \!\!\! f_i dx_i
= f_i^0 a_i^0 + f_i^1 a_i^1 + \ldots + f_i^n a_i^n,
\label{86}
\eeq
where $f_i^\ell$ is the value of $f_i$ at the $\ell$-th step, 
and $a_i^\ell$ is the increment in $x_i$ at the $\ell$-th step.
In this discretized form we see clearly that the path integral
depends on $\xi_0,\xi_1,\ldots,\xi_n$, and to find its
average we should use the path probability (\ref{82}). 
This amounts to multiply the right-hand side of (\ref{86}) by
the right-hand side of (\ref{82}) and integrate in all
variables, $\xi_0,\xi_1,\ldots,\xi_n$. The result of this
procedure is indicated by an index 'path' in the signs of the
average. Therefore, the average of $L$ which we call $W$ is
written as
\beq
W = \la L\ra_{\rm path}
\label{}
\eeq
Although we call work both $L$ and $W$, it should be understood
that $L$ is the actual work and $W$ its average.

The path integral (\ref{86a}) can be written as a time integral
of the power as is well known. A trajectory may be defined
parametrically by given $\xi_i$, and
thus $x_i$ and $p_i$, as functions of this parameter
which we take to be the time $t$. In terms of this
parameter, the integral (\ref{86a}) becomes a time integral,
\beq
L = \sum_i \int_{t_1}^{t_2} \!\!\! f_i \frac{p_i}m dt
\eeq
where we have replaced $dx_i/dt$ by the velocity $p_i/m$.
This expression is written as
\beq
L = \int_{t_1}^{t_2} \phi dt
\label{86b}
\eeq
where $\phi$ is the power at time $t$, and given by
\beq
\phi = \sum_i  f_i \frac{p_i}m
\label{86d}
\eeq
Taking the average of the expression (\ref{86b}) we find
\beq
\la L\ra_{\rm path} = \int_{t_1}^{t_2} \la \phi \ra dt
\label{86c}
\eeq
where in the right-hand side the average is the usual
average taken by the use of the probability density
$\rho(\xi,t)$ at time $t$ because $\phi$ depends only
on $\xi$ at time $t$. Thus the average over a path 
integral is transformed into an average over the ordinary
probability density. The integrand on the right-hand side
of (\ref{86c}) is understood as the average power, 
\beq
\Phi_w =\la \phi\ra = \int \sum_i  f_i \frac{p_i}m \rho d\xi
\eeq
which coincides with the power of the external force given by the
equivalent forms (\ref{67}) or (\ref{8w}), if we recall that
$f_i=-F_i^e$. Since $W$ is the average of $L$, we may write
\beq
W = \int_{t_1}^{t_2} \Phi_w dt
\label{92}
\eeq

The result (\ref{92}) or its equivalent form (\ref{86c})
where $L$ and $\phi$ are given by (\ref{86a}) and (\ref{86d}),
respectively, can immediately be generalized by replacing $f_i(\xi)$
by any other function of $\xi$. Let us suppose that
it is replaced by $J_i/\rho$ where $J_i$ is the irreversible
current given by (\ref{64})
\beq
J_i = D_i \rho - \frac{\Gamma_i}2 \frac{\partial \rho}{\partial p_i}
\eeq
The path integral of this quantity is
\beq
\Psi =  \sum_i \int_{\rm path} \frac{J_i}{\rho} dx_i
\eeq
and the associated flux according to the result above is
\beq
\Phi_q = \frac1m \sum_i \la \frac1{\rho }J_i p_i \ra
= \frac1m \sum_i \int J_i p_i d\xi
\eeq
which is the heat flux as given by (\ref{8x}).
The quantity 
\beq
Q = \int_{t_1}^{t_2} \Phi_q dt 
\eeq
is thus the heat exchanged between the two instants of time,
and according to the results above may be written as the
path average
\beq
Q = \la \Psi \ra_{\rm path}.
\label{99}
\eeq



Let us consider that a system in contact with a heat bath
at a temperature $T$ is acted by external forces during
a certain interval of time. The following equality relating the
free energy to the work performed by the system during
this interval of time has been shown to be valid
\cite{jarzynski1997},
\beq
e^{-\beta \Delta F} = \la e^{-\beta L} \ra_{\rm path}.
\label{42}
\eeq
Therefore, if the right-hand side of equation (\ref{42})
is measured, we may obtain $\Delta F$. 
Using the inequality $\la e^{\eta}\ra \geq e^{\la \eta \ra}$,
valid for any random variable $\eta$, we find
\beq
\Delta F \leq - \la L \ra_{\rm path} = - W,
\eeq
which should be compared with the equation (\ref{40}).


Before we end this section it is appropriate to place a discussion
on the experimental measurement of the several quantities
presented in the theory. The quantities that are mensurable
are those that we call state functions, that is, quantities that
are functions of the random variables, in the present
case the positions and momenta, and themselves random
variables. A experimental result obtained for a state function $E$,
be it the value of one trial or the arithmetic average
of several trials, should be compared with the average
predicted by the theory which is written as
\beq
\la E \ra = \int E(\xi)\rho(\xi) d\xi 
\eeq
such as the energy,
or as a path average as is the case of the mechanical work.

Let us consider the case of the entropy which is
\beq
S = -k \int \rho(\xi)\ln \rho(\xi) d\xi
\label{97} 
\eeq
which sometimes is written as
\beq
S = \la -k\ln \rho \ra
\label{97a}
\eeq
Although one may write in this form, this expression is merely
an abbreviation of the expression on the right-hand side of
(\ref{97}) and cannot be understood as the average of
a state function merely because $-k\ln\rho$ is not a
state function. Although, sometimes $-k\ln\rho$ is called
an instantaneous entropy, it is not a mensurable quantity.
This point can be better understood if we try to calculate
the entropy from a Monte Carlo simulation. One immediately
realizes that it is impossible to determine $\ln\rho$ 
along a Monte Carlo run and an alternative should be used.

The observation made above with respect to the average
(\ref{97a}) is extended to the average in (\ref{99})
because $J_i/\rho$ is not a state function and it is
not a mensurable quantity. This is paradoxical because
no one denies that heat $Q$ is mensurable. However,
a moment of reflection will reveal that heat is
measured through the work dissipated and not as
the quantity $\Psi$ above.

\section{Discussion and conclusion}

We have developed an approach to the stochastic thermodynamics
based on the use of the FPK equation, which governs the time
evolution of the probability distribution. 
The main feature of the approach in addition to the evolution
equation is the assignment of an energy function, the definition
of entropy and the introduction of an expression for the
rate of production of entropy. According to the
approach, these quantities are well defined
quantities in equilibrium or out of equilibrium.
This is in contrast to other quantities of thermodynamics
such as the temperature, which is defined only when
the system is in thermodynamic equilibrium. 

The evolution equation contains the mechanism 
of dissipation and stochastic fluctuation or noise
which leads the system toward equilibrium, if an
appropriate relation exists between dissipation
and noise. The mechanism is included in the
irreversible current by the term containing the
dissipative force and the term containing the quantity
$\Gamma$, which is a measure of the noise. 
If the thermodynamic equilibrium sets in, the
irreversible current vanishes. In out of equilibrium
the irreversible current is nonzero and the
production of entropy, which is related to the
square of the irreversible currents is greater than zero.
The rate of entropy production is thus a measure
of the deviation of a system from thermodynamic
equilibrium and of the irreversibility.

We have considered systems with a continuous space of states
in which case the appropriate evolution equation is the
FPK equation. However, the present approach can be extended
to a discrete space of states in which case the evolution
equation is called a master equation \cite{tome2015,tome2018}.
It can be applied to systems of interacting particles 
with different species including reactions among them \cite{tome2018}.
We did not treat the case where the parameters taking 
place in the evolution equation depends on time
but the present approach can also be applied for instance
to the case where the temperature
oscillate periodically in time  \cite{oliveira2019a,fiore2019}.
In this case, for long times the system may not properly reaches
a stationary state in the sense of being independent of time
but may reach a state with a probability that oscillates in time. 

Stochastic thermodynamics is sometimes called a discipline whose
quantities are defined at the level of single trajectories.
This denomination emphasizes the fluctuation aspect
of the theory, which is a relevant feature in systems
with few degrees of freedom, the main application of
the theory. In this respect the theory looks like
statistical mechanics, which incorporates fluctuations
and may also be applied to small systems. Thus, an
alternative name to discipline would be stochastic  
mechanics avoiding the term thermodynamics which
is usually associated to macroscopic systems. 

The emphasis on trajectories and path integralsl
is a distinguish feature as
is used for instance in the Jarzynski equality (\ref{42}) 
The present approach, on the other hand, the emphasis
rests on the fluxes and currents of various types
but a relationship between fluxes and path integrals 
exists as shown in section \ref{path}, revealing the
equivalence between the two approaches.
The present approach also emphasizes the 
connection with the laws of thermodynamics, particularly
the second law expressed by the nonnegativity of the
rate of entropy production.

The present approach to the quantum stochastic thermodynamics
is based on the quantum evolution equation which is a
canonical quantization of the FPK equation. It differs
from other approaches such as those based on the Lindblad
operators \cite{lindblad1976}. However, the present
quantum FPK equation has similarity with the quantum master equation
derived by Dekker \cite{dekker1977}
and by Caldeira and Leggett \cite{caldeira1983b,caldeira2014}. 
The main features of the quantum FPK equation
that distinguishes from other approaches is that it
is centered on the irreversible density current operator,
the analog of the classical irreversible probability current,
which plays a fundamental role in defining the fluxes
of various type as well as the rate of entropy production.
The similarity of the quantum evolution equation
to the classical counterpart   
is useful because the generalization of the concepts
of classical stochastic thermodynamics, such as those
associated to the current of probability, to the quantum
case become easier.
A final distinguishing feature is that a system described
by the quantum FPK equation thermalizes properly. That is,
for long times the system approaches the equilibrium state,
if, of course, the relationship (\ref{35}) between noise and
fluctuation is obeyed.

\appendix

\section{}
\label{A}

The FPK equation (\ref{2}) can be written in the form
\beq
\frac{\partial\rho}{\partial t} =
- \frac{\partial J_x}{\partial x} - \frac{\partial J_p}{\partial p},
\label{3a}
\eeq
where
$J_x$ and $J_p$ are the components of the probability current,
and given by
\beq
J_x = \frac{p}{m}, \qquad\qquad J_p = F\rho  
- \frac{\Gamma}2 \frac{\partial\rho}{\partial p},
\label{3b}
\eeq

Let us integrate both sides of the equation (\ref{3a}) in a region $R$
of the space $(x,p)$ delimited by a boundary line,
\beq
\frac{d}{dt}\int_R \rho \, dxdp
= - \int_R \frac{\partial J_x}{\partial x} dxdp
- \int_R \frac{\partial J_p}{\partial p} dxdp.
\eeq
The first integral can be integrated in $x$,
\beq
\int_R \frac{\partial J_x}{\partial x} dxdp
= \int [J_x(x_2,p)-J_x(x_1,p)]dp.
\eeq
For simplicity we are considering that $R$ is a convex region
so that there are two values of $x$ at the boundary of $R$ for
a given $p$, which we are denoting by $x_1(p)$ and $x_2(p)$.
In an analogous way we write the second integral as
\beq
\int_R \frac{\partial J_p}{\partial p} dxdp
= \int [J_p(x,p_1)-J_p(x,p_2)]dx.
\eeq

If $J_x$ and $J_p$ vanish at the boundary then both
integrals vanish and
\beq
\frac{d}{dt}\int_R \rho \, dxdp = 0,
\eeq
from which follows that the integral is a constant
that we set equal to unity,
\beq
\int_R \rho \, dxdp = 1.
\eeq
This result is extended to the case where the region $R$
is the whole space of states, in which case we demand that
$J_x$ and $J_p$ vanish at infinity, a requirement that is provided
by demanding that $\rho$ vanishes rapidly at infinity.

Let us consider now the case of an integral of the
type 
\beq
\int A \frac{\partial B}{\partial x} dxdp,
\eeq
where the integral is over the whole space of states.
If we perform and integration by parts the result is
\beq
\int \frac{\partial AB}{\partial x} dxdp 
- \int \frac{\partial A}{\partial x} B dxdp.
\eeq
Assuming that $AB$ vanishes rapidly at the limits of integration
as we did above, the first integral vanishes and we are left 
with the result
\beq
\int A \frac{\partial B}{\partial x} dxdp
= - \int \frac{\partial A}{\partial x} B dxdp.
\eeq

\section{}
\label{B}

Here we solve the FPK equation for the case of
a harmonic force $F^c=-Kx$. The equation is 
\beq
\frac{\partial\rho}{\partial t} =
- \frac{p}{m}\frac{\partial\rho}{\partial x}
+ K x \frac{\partial\rho}{\partial p}
+ \gamma \frac{\partial p\rho}{\partial p}  
+ \frac{m \gamma}\beta \frac{\partial^2\rho}{\partial p^2},
\label{2a}
\eeq
and it can be solved exactly by assuming the
following Gaussian form for the probability distribution
\beq
\rho = \frac1Z e^{-(ax^2+bp^2+2cxp)/2},
\label{55}
\eeq
where the parameters $a$, $b$, $c$ and $Z$ depend on time.
Replacing this form in the equation (\ref{2a}), we see
that the left and right-hand sides will only have terms
of the types $x^2$, $p^2$ and $xp$. Equating the respective
coefficients of these terms we find equations for the
parameters $a$, $b$, and $c$. There is no need to seek
an equation for $Z$ because this quantity can be obtained
from the three parameters $a$, $b$, and $c$. This follows
from the normalization of (\ref{55}), which gives 
\beq
Z=\int e^{-(ax^2+bp^2+2cxp)/2} dxdp = \frac{2\pi}{\sqrt{ab-c^2}}.
\eeq

Replacing the Gaussian distribution (\ref{55}) in the FPK equation
we may find the equations for the three parameters. However,
the equations are too complicated and we will instead seek
for equations that determine the covariances $\la x^2\ra$,
$\la p^2\ra$, and $\la xp\ra$. Before that we should write down
the relations between the covariances and the three
parameters, which are are obtained from the probability distribution
(\ref{55}), and are
\beq
\la x^2\ra
= \frac{b}{ab-c^2},
\eeq
\beq
\la p^2\ra
= \frac{a}{ab-c^2},
\eeq
\beq
\la xp\ra
= \frac{-c}{ab-c^2}.
\eeq
Inverting these relations we find
\beq
a = \frac{\la p^2\ra}{\la x^2\ra\la p^2\ra-\la xp\ra^2},
\eeq
\beq
b = \frac{\la x^2\ra}{\la x^2\ra\la p^2\ra-\la xp\ra^2},
\eeq
\beq
c = \frac{- \la xp\ra}{\la x^2\ra\la p^2\ra-\la xp\ra^2}.
\eeq

It remains now to determine the covariances as
functions of time. 
To find the equations for the covariances we proceed as follows.
We multiply both sides of the FPK equation successively by $x^2$,
$p^2$ and $xp$, and integrate in $x$ and $p$. Performing appropriate
integration by parts, we find
\beq
\frac{d}{dt}\la x^2\ra = \frac2m \la xp\ra,
\eeq
\beq
\frac{d}{dt}\la p^2\ra = -2 K \la xp\ra -2\gamma\la p^2\ra
+ \frac{2m\gamma}{\beta},
\eeq
\beq
\frac{d}{dt}\la xp\ra = \frac1m \la p^2\ra - K \la x^2\ra
-\gamma \la xp\ra.
\eeq

The stationary solution of this equation is $\la x^2\ra=1/K\beta$,
$\la p^2\ra=m/\beta$, and $\la xp\ra=0$. Taking these
result into account, we define variables that are deviations
of the covariances from their stationary values as follows,
$A=\la x^2\ra-1/K\beta$, $B=\la p^2\ra-m/\beta$, and $C=\la rp\ra$.
These variables obey the set of linear equations
\beq
\frac{dA}{dt} = \frac2m C,
\eeq
\beq
\frac{dB}{dt} = -2 K C -2\gamma B,
\eeq
\beq
\frac{dC}{dt} = \frac1m B - K A -\gamma C,
\eeq
which we write in matrix form
\beq
\frac{d}{dt}\left(
\begin{array}{c}
A \\
B \\
C \\
\end{array}
\right) = \left(
\begin{array}{rrr}
0  & 0        & 2/m \\
0  & -2\gamma & -2K \\
-K & 1/m      &  -\gamma\\
\end{array}
\right)\left(
\begin{array}{c}
A \\
B \\
C \\
\end{array}
\right).
\eeq

The solution for each variable is of the type $e^{\lambda t}$
where $\lambda$ is an eigenvalue of the square matrix above.
They are
\beq
\lambda_1 = -\gamma + \sqrt{\gamma^2 - 4K/m},
\eeq
\beq
\lambda_2 = -\gamma,
\eeq
\beq
\lambda_3 = -\gamma - \sqrt{\gamma^2 - 4K/m},
\eeq
and are all negative. The general solution is
\beq
A = A_1 e^{\lambda_1 t} + A_2 e^{\lambda_2 t}  + A_3 e^{\lambda_3 t},
\eeq
\beq
B = B_1 e^{\lambda_1 t} + B_2 e^{\lambda_2 t}  + B_3 e^{\lambda_3 t}, 
\eeq
\beq
C = C_1 e^{\lambda_1 t} + C_2 e^{\lambda_2 t}  + C_3 e^{\lambda_3 t},
\eeq
and the coefficients are not all independent, but are related by
\beq
\lambda_i A_i = \frac2m C_i,
\eeq
\beq
\lambda_i B_i = -2 K C_i -2\gamma B_i,
\eeq
\beq
\lambda_i C_i = \frac1m B_i - K A_i -\gamma C_i.
\eeq
Thus only three, say $A_1$, $A_2$, and $A_3$ can be chosen
to be independent, and they are determined by the initial
conditions.

It is worth determining the solution for long times.
In this case the solution is dominated by the largest
eigenvalue, which is $\lambda_1$. The covariances are 
\beq
\la x^2\ra = \frac1{K\beta} + A_1 e^{\lambda_1 t},
\eeq
\beq
\la p^2\ra = \frac{m}{\beta} + B_1 e^{\lambda_1 t},
\eeq
\beq
\la xp\ra = C_1 e^{\lambda_1 t},
\eeq
from which we find the three parameters
\beq
a = K \beta - a_1 e^{\lambda_1 t},
\eeq
\beq
b = \frac{\beta}m - b_1 e^{\lambda_1 t},
\eeq
\beq
c = - c_1 e^{\lambda_1 t},
\eeq
where $a_1=K^2\beta^2 A_1$, $b_1=\beta^2 B_1/m^2$, $c_1=K\beta^2 C_1/m$.

\section{}
\label{C}

We wish to determine here in an explicit form
the dissipative force $D=-\gamma$ for the quantum harmonic
oscillator were
\beq
g = - \frac{m}{i\hbar\beta}(e^{\beta H}xe^{-\beta H} - x),
\label{61}
\eeq
and
\beq
H = \frac1{2m}p^2 + \frac12 m\omega^2 x^2,
\eeq
To this end 
we start with the following identity \cite{merzbacher1970}
\[
e^{\beta H}xe^{-\beta H} = x + \beta [H,x] + \frac{\beta^2}{2}[H,[H,x]] + 
\]
\beq
+ \frac{\beta^3}{3!} [H,[H,[H,x]]] + \ldots
\eeq
Using the notation
\beq
A_n = [H,[H,\ldots [H,x]\ldots]],
\eeq
where the numbers of commutator is equal to $n$, 
the identity above is written as
\beq
e^{\beta H}xe^{-\beta H} = \sum_{n=0}^\infty \frac{\beta^n}{n!}A_n,
\label{60}
\eeq
where $A_0=x$. The quantities $A_n$ obeys the recursive relations
\beq
A_{n+1} = [H,A_n].
\eeq

To determine $A_n$ is easier if we use the relations
\beq
[H,x] = -\frac{i\hbar}{m} p,
\eeq
\beq
[H,p] = i m\omega^2\hbar x,
\eeq
which are
obtained by using the commutation relation $[x,p]=i\hbar$.
From these relations we get the two useful rules,
\beq
[H,[H,x] = \hbar^2 \omega^2 x,
\eeq
\beq
[H,[H,p] = \hbar^2 \omega^2 p.
\eeq

The first two coefficients of the expansion are
\beq
A_0 = x, \qquad\qquad
A_1 = -\frac{i\hbar}{m} p.
\eeq
Next, with the two rules above in mind,
we observe that $A_2$ will be proportional to $x$ and
$A_3$ will be proportional to $p$, and, in general, $A_n$
will be proportional to $x$ if $n$ is even, and it will be
proportional to $p$ if $n$ is odd. 

Let us consider the case $n$ even and write $A_n=a_n x$. Then
using the two rules above,
\beq
A_{n+2} = [H,[H,A_n]] = \hbar^2 \omega^2 a_n x,
\eeq
so that
\beq
a_{n+2} = \hbar^2\omega^2 a_n,
\eeq
from which we find
\beq
a_n = (\hbar\omega)^n,
\eeq
because $a_0=1$. The part of the expansion (\ref{60})
corresponding to $n$ even is
\beq
\sum_{n({\rm even})} \!\frac{\beta^n}{n!}A_n 
= x \!\!\sum_{n({\rm even})}\! \frac{(\beta\hbar\omega)^n}{n!} 
= x \cosh\beta\hbar\omega.
\eeq

Now we consider the case $n$ odd and write $A_n=b_n p$. Then
using the two rules above
\beq
A_{n+2} = [H,[H,A_n]] = \hbar^2 \omega^2 b_n p,
\eeq
so that
\beq
b_{n+2} = \hbar^2\omega^2 b_n,
\eeq
from which we find
\beq
b_n = -\frac{i}{m\omega}(\hbar\omega)^n,
\eeq
because $b_1=-i\hbar/m$. The part of the expansion (\ref{60})
corresponding to $n$ odd is
\beq
\sum_{n({\rm odd})} \!\frac{\beta^n}{n!}A_n 
= -\frac{i p}{m\omega} \sum_{n({\rm odd})}
\!\frac{(\beta\hbar\omega)^n}{n!}
= -\frac{i p}{m\omega} \sinh\beta\hbar\omega.
\eeq

Collecting the results above we find
\beq
e^{\beta H}xe^{-\beta H} = -\frac{i p}{m\omega} \sinh\beta\hbar\omega
+ x \cosh\beta\hbar\omega,
\label{60a}
\eeq
and the quantity (\ref{61}) becomes
\beq
g = p \frac{1}{\beta\hbar\omega} \sinh\beta\hbar\omega
+ i x \frac{m}{\beta\hbar}(\cosh\beta\hbar\omega -1).
\eeq

\section{}
\label{D}

Let suppose that $f$ is a function of several variables
that we denote by $x_i$, and that these 
variables depend on time. The derivative of $f$ with
respect to time is
\beq
\frac{df}{dt} = \sum_i f_i \frac{dx_i}{dt}
\label{79}
\eeq 
where $f_i$ are functions of $x_i$ given by
\beq
f_i = \frac{\partial f}{\partial x_i}
\eeq
Equation (\ref{79}) can be written in simplified form 
\beq
df = \sum_i f_i dx_i
\eeq
where $dx_i$ are the differentials of the variable $x_i$
and $df$ is the differential of $f$. Since $f$ is a
function of $x_i$ then the following relation is valid
\beq
\frac{\partial f_i}{\partial x_j} = \frac{\partial f_j}{\partial x_i}
\eeq

Now we raise the following question. Let $g$ be a function of $t$
and let $x_i$ depend on time as before and let us assume that
\beq
\frac{dg}{dt} = \sum_i g_i \frac{dx_i}{dt}
\label{80}
\eeq
where $g_i$ are given function of the variables $x_j$.
The question now arises whether $g$ could depend on time
only through the variables $x_i$, that is, whether
\beq
g(t) = g(x_1(t),x_2(t),\ldots)
\eeq
If that is possible then according to our reasoning above
the given functions $g_i$ of the variables $x_j$ must
fulfill the condition \beq
\frac{\partial g_i}{\partial x_j} = \frac{\partial g_j}{\partial x_i}
\eeq
for all pairs $i,j$. If this condition is not satisfied
it is not possible to write $g$ as a function of $x_i$. 
In this case if we write (\ref{80}) in the simplified form
\beq
dg = \sum_i g_i dx_i
\eeq
we say that $dg$ is not an exact differential.

\section{}
\label{E}

We determine here the covariances $\la x_ix_j\ra$, $\la x_ip_j\ra$
and $\la p_ip_j\ra$ for the system described by the FPK equation (\ref{76})
where$F_1^c=-Kx_1+Lx_2$, $F_2^c=-Kx_2+Lx_1$, and
$F_1^e= -c x_2$ and $F_2^e= c x_1$, 
which we reproduce here in the following form
\[
\frac{\partial\rho}{\partial t} =
-\frac{p_1}m \frac{\partial\rho}{\partial x_1}
-\frac{p_2}m \frac{\partial\rho}{\partial x_2}
\]
\[
+\frac{\partial}{\partial p_1}(Kx_1 + b x_2 +\gamma p_1)\rho
+\frac{\partial}{\partial p_2}(Kx_2 + a x_1 +\gamma p_2)\rho
\]
\beq
+ \gamma m k T_1 \frac{\partial^2\rho}{\partial p_1^2}
+ \gamma m k T_2 \frac{\partial^2\rho}{\partial p_2^2}
\label{76a}
\eeq
where $b =-L+c$ and $a =-L-c$.

Multiplying successively equation (\ref{76a}) by 
$x_ix_j$, $x_ip_j$, and $p_ip_j$, and performing the
integration we find the following equations, after
appropriate integration by parts,
\beq
\frac{d}{dt}\la x_1^2\ra = \frac2m\la x_1 p_1\ra
\eeq
\beq
\frac{d}{dt}\la x_2^2\ra = \frac2m\la x_2 p_2\ra
\eeq
\beq
\frac{d}{dt}\la x_1x_2\ra
= \frac1m\la x_1 p_2\ra + \frac1m\la x_2 p_1 \ra
\eeq

\beq
\frac{d}{dt}\la x_1p_2\ra = \frac1m\la p_1 p_2\ra - K\la x_1 x_2\ra 
- a \la x_1^2 \ra -\gamma\la x_1 p_2 \ra
\eeq
\beq
\frac{d}{dt}\la x_2p_1\ra = \frac1m\la p_1 p_2\ra - K\la x_1 x_2\ra 
- b \la x_2^2 \ra -\gamma\la x_2 p_1 \ra
\eeq
\beq
\frac{d}{dt}\la x_1p_1\ra = \frac1m\la p_1^2\ra - K\la x_1^2\ra 
- b \la x_1x_2 \ra -\gamma\la x_1 p_1 \ra
\eeq
\beq
\frac{d}{dt}\la x_2p_2\ra = \frac1m\la p_2^2\ra - K\la x_2^2\ra 
- a \la x_1x_2 \ra -\gamma\la x_2 p_2 \ra
\eeq

\beq
\frac{d}{dt}\la p_1^2\ra = -2K\la x_1p_1\ra - 2 b \la x_2p_1\ra
-2\gamma \la p_1^2\ra + 2\gamma mkT_1 
\eeq
\beq
\frac{d}{dt}\la p_2^2\ra = -2K\la x_2p_2\ra - 2 a \la x_1p_2\ra
-2\gamma \la p_2^2\ra + 2\gamma mkT_2 
\eeq
\[
\frac{d}{dt}\la p_1p_2\ra = -K\la x_1p_2\ra - K\la x_2p_1\ra 
\]
\beq
- b \la x_2p_2\ra - a \la x_1p_1\ra -2\gamma \la p_1p_2\ra 
\eeq

Now we look for the stationary solution. Setting the above
equation to zero, we find that the following covariances
vanish, $\la x_1p_1\ra=0$, $\la x_2 p_2\ra = 0$, $\la p_1p_2\ra = 0$.
The other covariances are the solution of the set of linear
equations
\beq
\la x_1 p_2\ra + \la x_2 p_1 \ra = 0
\eeq
\beq
K\la x_1 x_2\ra + a \la x_1^2 \ra + \gamma\la x_1 p_2 \ra = 0
\eeq
\beq
K\la x_1 x_2\ra + b \la x_2^2 \ra + \gamma\la x_2 p_1 \ra = 0
\eeq
\beq
\la p_1^2\ra - mK\la x_1^2\ra - b m\la x_1x_2 \ra = 0
\eeq
\beq
\la p_2^2\ra - mK\la x_2^2\ra - a m\la x_1x_2 \ra = 0
\eeq
\beq
b \la x_2p_1\ra + \gamma \la p_1^2\ra = \gamma mkT_1
\eeq
\beq
a \la x_1p_2\ra + \gamma \la p_2^2\ra = \gamma mkT_2
\eeq
A straightforward calculation leads us to the result
\beq
\la x_1p_2\ra = - \la x_2p_1\ra = 
\frac{m\gamma k(bT_2-aT_1)}{2(m\gamma^2 K + ab)}
\eeq
\beq
\la x_1 x_2 \ra = - \frac{k(aT_1+bT_2)}{2(K^2-ab)}
\eeq
\beq
\la p_1^2\ra = m k T_1 + \frac{b}{\gamma} \la x_1p_2\ra
\eeq
\beq
\la p_2^2\ra = m k T_2 - \frac{a}{\gamma} \la x_1p_2\ra
\eeq
\beq
\la x_1^2\ra = - \frac{\gamma}a \la x_1p_2\ra - \frac{K}a \la x_1x_2\ra
\eeq
\beq
\la x_2^2\ra = \frac{\gamma}b \la x_1p_2\ra - \frac{K}b \la x_1x_2\ra
\eeq
We remark that, as $\la x_1^2\ra$, $\la x_2^2\ra$, $\la p_1^2\ra$, and
$\la p_2^2\ra$ must be nonnegative,
the following conditions should be fulfilled
\beq
mK\gamma^2 + ab \geq 0,
\qquad
K^2-ab\geq 0.
\eeq

It is worth mentioning that the probability density
can also be determined. On account of the linearity
of the FPK equation in relation to the variable
$x_i$ and $p_i$, the solution is a multivariate
Gaussian distribution, which we write as 
\beq
\rho = \frac1Z \exp\{ - \frac12 \sum_{i,j=1}^4L_{ij}\xi_i\xi_j\}
\label{88}
\eeq
where we are using the abbreviations $\xi_1=x_1$,
$\xi_2=x_2$, $\xi_3=p_1$, and $\xi_4=p_2$. 
The matrix $L$ whose elements are $L_{ij}$ is the
inverse of the covariance matrix $C$, whose elements
we have just determined, and are $C_{11}=\la x_1^2\ra$,
$C_{22}=\la x_2^2\ra$, $C_{12}=\la x_1x_2\ra$,
$C_{33}=\la p_1^2\ra$, $C_{44}=\la p_2^2\ra$,
$C_{34}=\la p_1p_2\ra$,$C_{13}=\la x_1p_1\ra$,
$C_{14}=\la x_1p_2\ra$, $C_{23}=\la x_2p_1\ra$,
and $C_{24}=\la x_2p_2\ra$. The expression given by equation
(\ref{88}) is the probability distribution describing
the nonequilibrium stationary state of the present problem.


\end{document}